\documentclass[superscriptaddress,showpacs,aps,twocolumn,floatfix,prb,longbibliography]{revtex4-2}
\usepackage{bm,amsmath,amssymb, mathrsfs}
\usepackage{amssymb}
\usepackage{commath}
\usepackage{braket} 
\usepackage{graphicx,bm}
\usepackage{verbatim}
\usepackage[dvipsnames]{xcolor}
\usepackage{soul}
\usepackage[colorlinks]{hyperref}
\AtBeginDocument{%
    \hypersetup{    
        linkcolor=blue,
        filecolor=magenta,      
        urlcolor=blue,
        citecolor=blue,
        colorlinks=true,
     }
}

\newcommand{\eu}{\mathrm{e}}
\newcommand{\imi}{\mathrm{i}}
\newcommand{\der}{\mathrm{d}}

\usepackage[T1]{fontenc}
\usepackage[utf8]{inputenc}

\begin{document}

\title{Transport signatures of inverted Andreev bands in topological Josephson junctions}

\author{Jonathan Sturm}
\email{jonathan.sturm@uni-wuerzburg.de}
\affiliation{Institute for Theoretical Physics and Astrophysics and W{\"u}rzburg-Dresden Cluster of Excellence ct.qmat,\\
	Julius-Maximilians-Universit{\"a}t W{\"u}rzburg, 97074 W{\"u}rzburg, Germany}

\author{Raffael L. Klees}
\affiliation{Institute for Theoretical Physics and Astrophysics and W{\"u}rzburg-Dresden Cluster of Excellence ct.qmat,\\
	Julius-Maximilians-Universit{\"a}t W{\"u}rzburg, 97074 W{\"u}rzburg, Germany}
\affiliation{Institute of Physics, University of Augsburg, D-86159 Augsburg, Germany}

\author{Ewelina M. Hankiewicz}
\affiliation{Institute for Theoretical Physics and Astrophysics and W{\"u}rzburg-Dresden Cluster of Excellence ct.qmat,\\
	Julius-Maximilians-Universit{\"a}t W{\"u}rzburg, 97074 W{\"u}rzburg, Germany}

\author{Daniel Gresta}
\affiliation{Institute for Theoretical Physics and Astrophysics and W{\"u}rzburg-Dresden Cluster of Excellence ct.qmat,\\
	Julius-Maximilians-Universit{\"a}t W{\"u}rzburg, 97074 W{\"u}rzburg, Germany}

\date{\today}

\begin{abstract}
We study the thermoelectrical transport transverse to conventional and topological Josephson junctions with a central quantum dot (QD).
For that purpose, we derive an effective resonant tunneling model where the QD is renormalized with an induced superconducting gap. 
By applying the Keldysh Green's function technique, we compute the local density of states as well as the transmission functions. 
In the latter case, we observe that the Andreev bound states forming on the QD are inverted if the junction has $p$-wave symmetry, meaning that electron and hole orbitals switch roles. 
We calculate the thermoelectric transport coefficients both analytically and numerically and show how the induced gaps and the band inversion are reflected in the electrical and heat conductance as well as the Seebeck coefficient, the latter experiencing a sign change in the topological case.
\end{abstract}

\maketitle

\section{\label{sec: Introduction} Introduction}
Two decades ago, the seminal work by Kitaev \cite{draft_Kitaev2001} predicted the presence of Majorana zero modes (MZMs) at the boundaries of spinless superconductors (SCs) with $p$-wave pairing symmetry. However, the experimental verification of their existence is still an open problem.
Although it is a well-established fact that coupling a metal electrode to a MZM leads to a zero voltage conductance peak \cite{draft_Law2009, draft_Flensberg2010, draft_Sau2010, draft_Mourik2012}, it has been shown that such a peak is not an exclusive feature of MZMs but can be generated by any zero-energy mode \cite{draft_Moore2018, draft_Vuik2019, draft_Yu2021, draft_Klees2023}.

By coupling two $p$-wave SCs to each other, a topological Josephson junction (JJ) is formed.
The interference of the two Majorana fermions induces a MZM in the Andreev bound states (ABSs) at a phase difference of $\pi$, leading to a $4\pi$-periodic Josephson current \cite{draft_Fu2008, draft_Tkachov2013, draft_Peng2016, draft_Deacon2017}.
However, also this MZM signature turned out to be ambiguous as it can be mimicked by topologically trivial JJs \cite{draft_Zazunov2018, draft_Chiu2019}.

In search of alternative ways to prove the existence of MZMs in topological SC structures, thermal and thermoelectric transport have proven themselves powerful tools.
For heat transport, it has been shown that the thermal conductance along a topological JJ is quantized \cite{draft_Fu2008, draft_Sothmann2016} and that the $4\pi$-periodic ground-state fermion parity can be measured in phase-dependent thermodynamics \cite{draft_Scharf2020}.
Moreover, several studies have found a sign change in the Seebeck coefficient of a resonant tunneling model (RTM), that is, a normal metal-quantum dot-normal metal structure, when the quantum dot (QD) is coupled to a MZM \cite{draft_Lopez2014, draft_Valentini2015, draft_Grosu2023}.
However, so far no intuitive physical explanation for this anomalous Seebeck effect has been given.

In this work, we address this problem by tracing back the sign change to a band inversion on the QD by showing that electron and hole orbitals exchange roles.
For that, we extend the RTM by coupling two additional SC leads to the QD, obtaining the four-terminal structure shown in Fig.~\ref{fig draft 1}~(a).
The two SCs, which will be of $s$- or $p$-wave type, induce ABSs on the QD, which can be probed by transverse transport between the normal leads. 
Such a geometry has been successfully used in a previous work \cite{draft_Bauer2021} to theoretically show that the transverse heat conductance in a topological junction is quantized.

This paper is organized as follows:
In Sec. II, we set the theoretical stage by stating the mathematical model and explaining the nonequilibrium Green's function formalism used to compute the transport observables.
In Sec. III, we analyze the local density of states and the transmission functions of the different transport processes to reveal the band inversion. 
We show how the band inversion can be measured in the conductance as well as the Seebeck coefficient in Sec. IV. 
Finally, in Sec. V we give a short summary of our findings and an outlook to future directions in this field.

\section{\label{sec: Setup} Setup and Formalism}

\subsection{\label{sec: Model} Model}

The system under consideration is a four-terminal junction consisting of two SC leads (S) with labels $\{1,2\}$ and two normal metallic electrodes (N)  labeled by $\{\mathrm{L,R}\}$ that are all coupled to a central QD with energy level $\varepsilon_0$; cf. Fig.~\ref{fig draft 1}~(a). 
With that, an S-QD-S Josephson junction is formed in the vertical direction, while on the horizontal we have an N-QD-N resonant tunneling model.
Both systems separately have been studied extensively in the literature; see, e.g., Refs.~\cite{draft_Yeyati2011} and \cite{draft_Cuevas}, respectively.
The coupling to the S leads induces ABSs on the QD, which depend on the phase difference $\phi=\varphi_1-\varphi_2$ between the SCs and can be modified with the energy level of the QD $\varepsilon_0$.
These ABSs in turn define an effective RTM with a superconducting QD [cf. Fig.~\ref{fig draft 1}~(b)] that can be probed by means of thermoelectric transport between the normal leads, which is the central goal of this article.

The Hamiltonian describing the full system is given by
\begin{equation}\label{eq draft Hamiltonian}
    H = H_{\rm S}^\alpha + H_{\rm N} + H_{\rm T} + H_0 \; .
\end{equation}
$H_{\rm S}^\alpha=H_1^\alpha+H_2^\alpha$ is the Hamiltonian of the SC leads 1 and 2. 
The index $\alpha \in \{ s,p \}$ indicates the pairing symmetry of the SC leads.
If $\alpha=s$, both leads are of $s$-wave symmetry, forming a conventional JJ, while for $\alpha=p$, $H_{1,2}^p$ describe a pair of semi-infinite Kitaev chains \cite{draft_Kitaev2001, draft_zazunov2016}, leading to a topological JJ.
Therefore, in the case $\alpha=p$ we consider the whole system to be fully spin-polarized.
$H_{\rm N}=H_{\rm L}+H_{\rm R}$ models the normal metallic leads L,R. 
The tunneling between the QD and the leads is given by the Hamiltonian
\begin{equation}
    H_{\rm T} = \sum_{i=1,2,\rm{L,R}} \sum_{\sigma} t_i \left( c_{0\sigma}^\dagger c_{i\sigma} + c_{i\sigma}^\dagger c_{0\sigma}\right) \; ,
\end{equation}
with hopping amplitudes $t_i\in\mathbb{R}$.
Here, $c_{i\sigma}^{(\dagger)}$ and $c_{0\sigma}^{(\dagger)}$ are the fermionic operators of lead $i$ and the QD, respectively. 
In the case $\alpha=s$, the spin index takes values $\sigma=  {\uparrow,\downarrow}$, while for $\alpha=p$ we choose the polarization as $\sigma= \ \uparrow$.
Lastly, the QD is assumed to be spin-degenerate and non-interacting with a single energy level $\varepsilon_0$, yielding the Hamiltonian 
\begin{equation}
    H_0 = \sum_\sigma \varepsilon_0 c_{0\sigma}^\dagger c_{0\sigma} \; .
\end{equation}

The physics of the boundaries of the leads that are connected to the QD are fully contained in their boundary Green's functions (GFs).
We use the Nambu basis spinors $\psi_i^\dagger = (c_{i\uparrow}^\dagger, c_{i\downarrow})$ ($i\in\{1,2,\mathrm{L,R},0\}$), whereby the spin index is only present in the case $\alpha=s$.
With that, the boundary GFs for the S leads in the wide-band limit read \cite{draft_zazunov2016}
\begin{subequations}\label{eq GF super}
\begin{align}
    g_{1,2}^s (\varepsilon) & = -\pi\nu_0 \frac{\varepsilon \tau_0 + \Delta \eu^{\imi\varphi_{1,2}\tau_3}\tau_1}{\sqrt{\Delta^2-\varepsilon^2}} \; , \\
    g_{1,2}^p (\varepsilon) & =  \frac{\pi\nu_0}{\varepsilon}  \left(\sqrt{\Delta^2-\varepsilon^2}\tau_0 \pm \Delta\eu^{\imi\varphi_{1,2}\tau_3}\tau_1 \right) \; .
\end{align}
\end{subequations}
Here, $\Delta > 0$ is the SC gap, $\varphi_{1,2}$ are the SC phases, and $\nu_0$ is the density of states at the Fermi level in the normal-conducting state. 
$\tau_{j=0,\dots,3}$ are the identity and Pauli matrices in Nambu space.
The sign change of $\Delta$ between $g_{1}^p$  and $g_2^p$ is a direct consequence of the antisymmetry of the $p$-wave pairing \cite{draft_zazunov2016}, while the $s$-wave pairing is symmetric and, therefore, there is no sign change between $g_1^s$ and $g_2^s$.
We will show how this can be used to measure the pairing symmetry and characterize the junction below.
The corresponding retarded and advanced GFs, $g_{1,2}^{\alpha,\rm{r/a}}$, 
which are implicitly used throughout the following calculations,
are obtained by shifting $\varepsilon \mapsto \varepsilon\pm \imi\eta$, where $\eta>0$ is the Dynes parameter \cite{draft_Dynes1978}, which phenomenologically accounts for small disorder.
The GFs for the normal leads then follow by taking the limit $\Delta\to0$ \cite{draft_Cuevas, draft_Klees2023}:
\begin{equation}
    g_{\rm L,R}^{\rm r/a} = \lim_{\Delta\to0} g_{1,2}^{\alpha, \rm r/a} (\varepsilon) = \mp \imi \pi\nu_0 \tau_0 \; .
\end{equation}
Finally, the couplings in Nambu space read $V_{i0}=V_{0i}=t_i \tau_3$ and the Hamiltonian of the QD is $\mathcal{H}_0= \varepsilon_0 \tau_3$.
Note that for analytical purposes, the limit $\eta\to0$ is implicitly taken at the end of all calculations, while the numerical calculations use $\eta=0.001\Delta$.

\subsection{Self-energies and induced gaps}

\begin{figure}
    \centering
    \includegraphics[width=\linewidth]{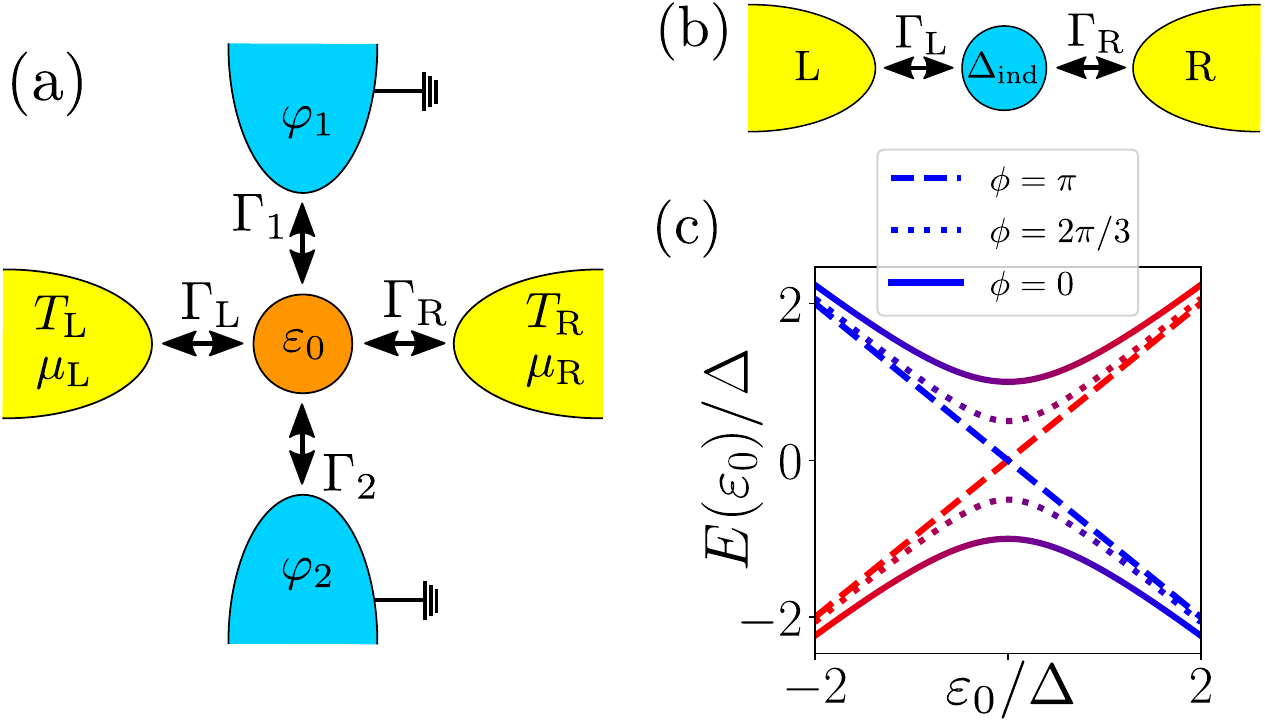}
    \caption{(a) Two SCs (blue) with phases $\varphi_{1,2}$ and two normal leads (yellow) with chemical potentials $\mu_{\rm L,R}$ and temperatures $T_{\rm L,R}$ are coupled to a QD with energy level $\varepsilon_0$ (orange) via couplings $\Gamma_{1,2}$ and $\Gamma_{\rm L,R}$.
    The SCs are grounded, implying $\mu_{1,2} \equiv 0$.
    (b) Effective RTM with a superconducting QD with induced pairing $\Delta_{\rm ind}$.
    (c) Quasiparticle excitation spectrum $E = \pm\sqrt{\varepsilon_0^2 + |\tilde{\Delta}_{\rm ind}^s|^2}$, cf. Eq.~\eqref{eq draft effective Hamiltonian} and the discussion thereunder, for various phase differences $\phi=\varphi_1-\varphi_2$.
    The red (blue) segments indicate electron-like (hole-like) quasiparticles.
    The particle hole mixing is indicated by the color gradient.
    We used $\Gamma_1=\Gamma_2=\Delta/2$.}
    \label{fig draft 1}
\end{figure}
For the transport formalism that we will make use of below, we need the dressed GF of the QD, $G_0^\alpha$, which can be found from the Dyson equation as 
\begin{equation}\label{eq draft dressed GF}
    G_0^\alpha = (\varepsilon - \mathcal{H}_0 - \Sigma_{\rm S}^\alpha - \Sigma_{\rm N})^{-1} \; .
\end{equation}
The self-energy originating from the normal leads is given by 
\begin{equation}
    \Sigma_{\rm N} = \sum_{i=\rm L,R} V_{0i} g_i^{\rm r} V_{i0} = - \imi (\Gamma_{\rm L}+\Gamma_{\rm R})\tau_0 \; ,
\end{equation}
where we defined $\Gamma_i = \pi\nu_0 t_i^2$.
In the same way one can obtain the self-energies from the coupling to the SCs, $\Sigma_{\rm S}^\alpha = \sum_{i=1,2} V_{0i} g_i^\alpha V_{i0}$, whose explicit forms are
\begin{subequations}\label{eq draft self-energies}
\begin{align}
    \Sigma_{\rm S}^s(\varepsilon) & = 
    \begin{pmatrix}
        -\tilde{\varepsilon}(\Gamma_1+\Gamma_2) & \Delta_{\rm ind}^s \\
        (\Delta_{\rm ind}^s)^* & -\tilde{\varepsilon}(\Gamma_1+\Gamma_2)
    \end{pmatrix} \; , \\
    \label{eq draft self-energies b}
    \Sigma_{\rm S}^p(\varepsilon) & = 
    \begin{pmatrix}
        \tilde{\varepsilon}^{-1}(\Gamma_1+\Gamma_2) & \Delta_{\rm ind}^p \\
        (\Delta_{\rm ind}^p)^* & \tilde{\varepsilon}^{-1}(\Gamma_1+\Gamma_2)
    \end{pmatrix} \; ,
\end{align}
\end{subequations}
where $\tilde{\varepsilon}=\frac{\varepsilon}{\sqrt{\Delta^2-\varepsilon^2}}$.
As the off-diagonal terms $\Delta_{\rm ind}^\alpha$ induce mixing of particles and holes, we call them induced gaps, which have the form 
\begin{subequations}\label{eq draft induced}
\begin{align}
    \Delta_{\rm ind}^s & = \frac{\Delta}{\sqrt{\Delta^2-\varepsilon^2}} (\Gamma_1\eu^{\imi\varphi_1}+\Gamma_2\eu^{\imi\varphi_2}) \label{eq draft induced s}  , \\
    \Delta_{\rm ind}^p & = -\frac{\Delta}{\varepsilon} (\Gamma_1\eu^{\imi\varphi_1} - \Gamma_2\eu^{\imi\varphi_2}) \label{eq draft induced p} .
\end{align}
\end{subequations}
Notice that while the induced $s$-wave gap $\Delta_{\rm ind}^s$ is symmetric under exchange of the two SCs, $\Delta_{\rm ind}^p$ is antisymmetric.
This property is inherited from the signs of the pairing terms in the GFs in Eq.~\eqref{eq GF super} and hence, as pointed out above, a direct consequence of the different pairing symmetries.  
To make this property more evident, we consider the absolute values of $\Delta_{\rm ind}^\alpha$, yielding
\begin{subequations}
\begin{align}
    |\Delta_{\rm ind}^s|^2 & \propto (\Gamma_1-\Gamma_2)^2 + 4\Gamma_1\Gamma_2\cos^2(\phi/2) \; , \\
    |\Delta_{\rm ind}^p|^2 & \propto (\Gamma_1-\Gamma_2)^2 + 4\Gamma_1\Gamma_2\sin^2(\phi/2) \; .
\end{align}
\end{subequations}
Therefore, $|\Delta_{\rm ind}^s|$ ($|\Delta_{\rm ind}^p|$) is minimal (maximal) for $\phi=\pi$ ($\phi=0)$, while being maximal (minimal) for $\phi=0$ ($\phi=\pi$).
The strong dependence of the induced gap on the phase difference, which originates from the interference of the ground-state wavefunctions of the SC leads, can be used to control the mixing of particles and holes on the QD. 
To illustrate this, consider the $s$-wave case in the low-energy limit $\varepsilon/\Delta\to0$, allowing us to define an effective Hamiltonian $\mathcal{H}_{\rm eff}$ of the QD as
\begin{equation}\label{eq draft effective Hamiltonian}
    \mathcal{H}_{\rm eff} = \mathcal{H}_0 + \Sigma_{\rm S}^s(0) = 
    \begin{pmatrix}
        \varepsilon_0 & \tilde{\Delta}_{\rm ind}^s \\
        (\tilde{\Delta}_{\rm ind}^s)^* & -\varepsilon_0
    \end{pmatrix} \; ,
\end{equation}
where now $\tilde{\Delta}_{\rm ind}^s = \Gamma_1\eu^{\imi\varphi_1}+\Gamma_2\eu^{\imi\varphi_2}$.
The spectrum of this Hamiltonian is then given by $E = \pm\sqrt{\varepsilon_0^2 + |\tilde{\Delta}_{\rm ind}^s|^2}$, which is shown in Fig.~\ref{fig draft 1}~(c) for the case $\Gamma_1=\Gamma_2$.
As one can see, the quasiparticle excitation spectrum on the QD (i.e., the ABSs) as a function of $\varepsilon_0$ resembles the common spectrum of an $s$-wave SC with the additional parameter $\phi$ enabling opening and closing the gap. 
With that, the phase difference allows us to continuously transform the effective N-S-N-junction in Fig.~\ref{fig draft 1}~(b) ($\phi\neq\pi$) into a standard RTM ($\phi=0$).
Note that if $\Gamma_1\neq\Gamma_2$, this statement still holds approximately if the deviation is small ($|\Gamma_1-\Gamma_2|/\Gamma_1 \ll 1$), which is the case we are interested in.
This will help us to interpret the ABSs and transmission functions below.

\begin{figure*}
    \centering
    \includegraphics[width=\linewidth]{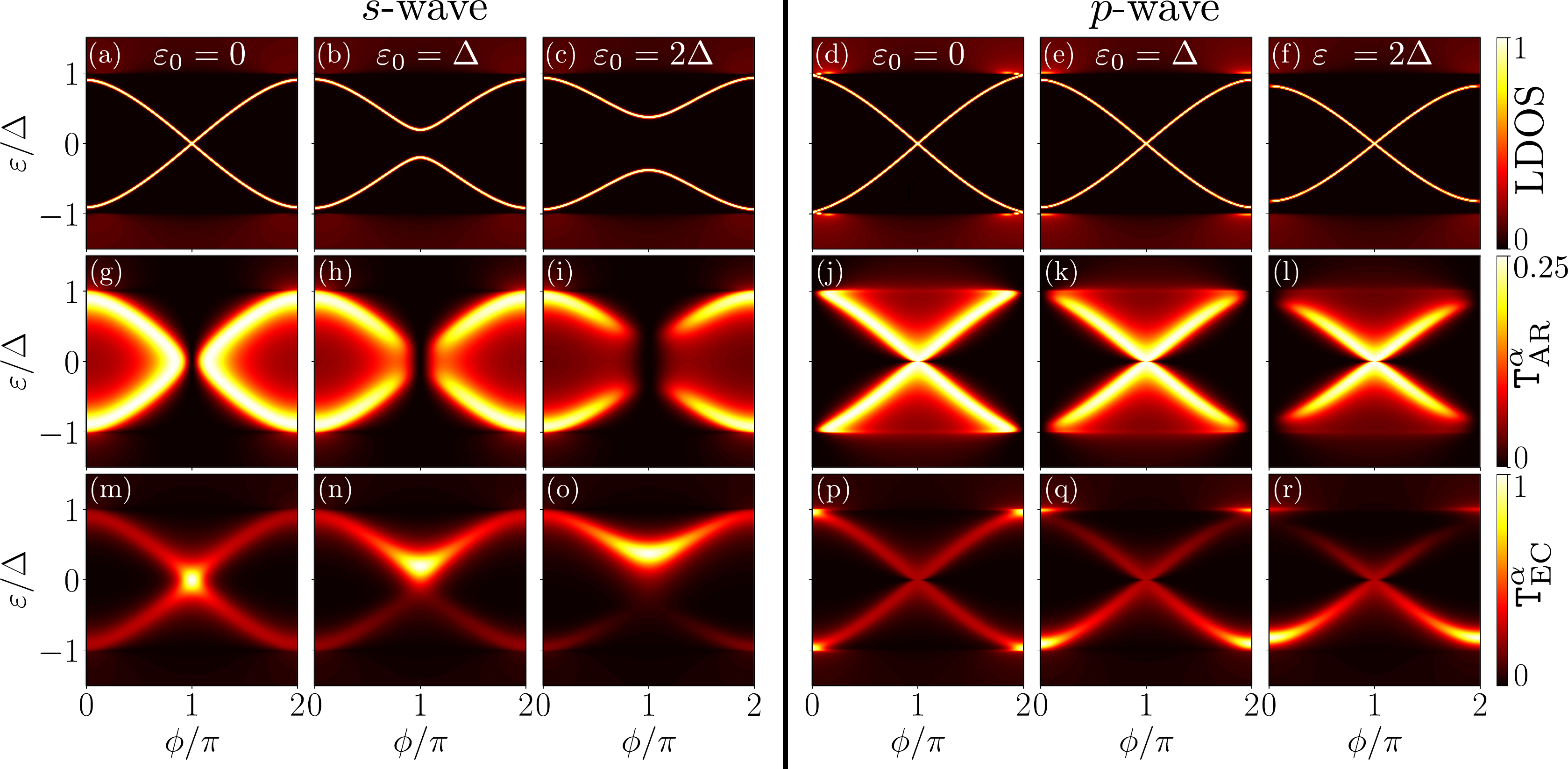}
    \caption{First row: LDOS on the QD, cf. Eq.~\eqref{eq draft LDOS}, as a function of $\phi$ and $\varepsilon$.
    We used $\Gamma_1=\Gamma_2=2\Delta$ and $\Gamma_{\rm L}=\Gamma_{\rm R}=0$.
    Second and third rows: $\mathtt{T}_{\rm AR}^\alpha$ and $\mathtt{T}_{\rm EC}^\alpha$, respectively.
    Here we used $\Gamma_1=\Gamma_2=2\Delta$ and $\Gamma_{\rm L}=\Gamma_{\rm R}=\Delta/2$.
    In all plots we used the Dynes parameter $\eta=0.001\Delta$.}
    \label{fig draft 2}
\end{figure*}
\subsection{Transport formalism}
Different from many transport studies of JJs, we are interested in the thermoelectric transport between the two normal leads rather than the Josephson current along the JJ.
This approach has several advantages:
First, we can study the electrical conductance without applying a voltage bias between the SCs and avoid driving the system in a fully nonequilibrium state \cite{draft_cuevas1996}. 
Second, we can study thermoelectric effects by heating the metallic leads rather than the SCs, which is favorable from the experimental point of view.
Third, the transverse approach allows us to directly study the subgap physics rather than the quasiparticle transport above the gap.

To do so, we define the electrical and heat currents in the left lead by means of the Heisenberg equation of motion as $I_{\rm L} = \frac{e}{\imi\hbar}\braket{[ H,N_{\rm L} ]}$ and $J_{\rm L}=-\frac{1}{\imi\hbar}\braket{[H, H_{\rm L} - \mu_{\rm L}N_{\rm L}]}$, respectively, where $e>0$ is the elementary charge, $\hbar$ is the reduced Planck constant, and $N_{\rm L}=\sum_\sigma c_{\rm L\sigma}^\dagger c_{\rm L\sigma}$ is the particle number operator of the left lead.
Let $\mu_{\rm L,R}$ and $T_{\rm L,R}$ be the chemical potentials and temperatures of the left and right leads, respectively.
Since we assume the SCs to be grounded, we set their chemical potentials $\mu_{1,2}\equiv0$.
Moreover, we set $\mu_{\rm L}=\mu+\Delta\mu$ and $\mu_{\rm R}=\mu$, as well as $T_{\rm L}=T+\Delta T$ and $T_{\rm R}=T$.
In a linear response regime, where $\Delta\mu\ll k_{\rm B}T$ and $\Delta T\ll T$ are small compared to the reference temperature $T$ ($k_{\rm B}$ is the Boltzmann constant), the currents are related to the biases by means of the Onsager matrix (cf. Appendix~\ref{sec draft app B}) \cite{draft_benenti2017}: 
\begin{equation}\label{eq draft onsager}
    \begin{pmatrix}
        I_{\rm L}/e \\
        J_{\rm L}/(k_{\rm B}T)
    \end{pmatrix}
    = \frac{1}{h}
    \begin{pmatrix}
        L_{11}^\alpha & L_{12}^\alpha \\ L_{21}^\alpha & L_{22}^\alpha
    \end{pmatrix}
    \begin{pmatrix}
        \Delta\mu \\
        k_{\rm B} \Delta T
    \end{pmatrix},
\end{equation}
where $L^\alpha=(L_{ij}^\alpha)_{i,j=1,2}$ is the Onsager matrix whose coefficients are given by
\begin{equation}\label{eq draft onsager coefficients}
    L_{ij}^\alpha = \int_{-\infty}^\infty 
    \left(
    \frac{\varepsilon-\mu}{k_{\rm B}T}
    \right)^{i+j-2}
    \mathtt{T}_{ij}^\alpha(\varepsilon) 
    \left(
    -\frac{\partial f}{\partial\varepsilon}
    \right)
    \der\varepsilon,
\end{equation}
with the Fermi function $f(\varepsilon)=(1 + \exp(\frac{\varepsilon-\mu}{k_{\rm B}T}))^{-1}$.
As we show in Appendix~\ref{sec draft app A}, the transmission functions are given by 
\begin{subequations} \label{eq draft transmission onsager}
    \begin{align}
        \mathtt{T}_{11}^\alpha = \mathtt{T}_{21}^\alpha & = \mathtt{T}_{\rm EC}^\alpha + \mathtt{T}_{\rm CAR}^\alpha + 2\mathtt{T}_{\rm AR}^\alpha, \\
        \mathtt{T}_{12}^\alpha = \mathtt{T}_{22}^\alpha & = \mathtt{T}_{\rm EC}^\alpha + \mathtt{T}_{\rm CAR}^\alpha,
    \end{align}
\end{subequations}
satifsying the Fisher-Lee relations \cite{draft_Fisher1981}
\begin{subequations}\label{eq draft transmission}
\begin{align}
    \mathtt{T}_{\rm AR}^\alpha & = 4\Gamma_{\rm L}^2 \big| G_0^{\rm \alpha, eh}\big|^2 \; ,\\
    \mathtt{T}_{\rm EC}^\alpha & = 4\Gamma_{\rm L}\Gamma_{\rm R} \big| G_0^{\rm \alpha,ee} \big|^2 \; , \label{eq draft TEC} \\
    \mathtt{T}_{\rm CAR}^\alpha & = 4\Gamma_{\rm L}\Gamma_{\rm R} \big| G_0^{\rm \alpha, eh} \big|^2 \; .
\end{align}
\end{subequations}
The three transmission functions describe the probabilities of three transport processes:
$\mathtt{T}_{\rm AR}^\alpha$ is the probability of Andreev reflection (AR), i.e., converting an electron from the left lead into a hole in the same lead.
The elastic cotunneling (EC) probability is given by $\mathtt{T}_{\rm Ec}^\alpha$ and is the usual tunneling of an electron from the left to the right lead via the QD.
Lastly, crossed Andreev reflection (CAR) is the process of an electron from the left lead being converted into a hole in the right lead with probability $\mathtt{T}_{\rm CAR}^\alpha$.
Note that we only consider subgap transport processes since we are only interested in small biases around the Fermi level $\mu=0$ and, therefore, can neglect quasiparticle transport above the gap.

There are four different thermoelectric phenomena contained in Eq.~\eqref{eq draft onsager}:
The most common one is the generation of an electrical current under a finite bias $\Delta\mu\neq0$ while $\Delta T=0$, which is described by the electrical conductance $G^\alpha$.
Under the same conditions also a heat current is flowing, which is known as the Peltier effect with Peltier coefficient $\Pi^\alpha$.
On the other hand, also a finite temperature gradient $\Delta T\neq0$ will generate a heat current, where the heat conductance $K^\alpha$ is defined for an electrically insulating system ($I_{\rm L}=0$).
In this case, a voltage bias forms to balance out the temperature bias such that there is no net charge transport.
The formation of this voltage bias is known as the Seebeck effect with Seebeck coefficient (of thermopower) $S^\alpha$.

In terms of the Onsager matrix elements, the four transport coefficients are given by
\begin{subequations}\label{eq draft transport coefficients}
    \begin{align}
        G^\alpha & = G_0 L_{11}^\alpha, \\
        S^\alpha & = \frac{k_{\rm B}}{e} \frac{L_{12}^\alpha}{L_{11}^\alpha}, \\
        \Pi^\alpha & = \frac{k_{\rm B}T}{e} \frac{L_{21}^\alpha}{L_{11}^\alpha}, \\
        K^\alpha & = \frac{3K_0}{\pi^2} \frac{\det(L^\alpha)}{L_{11}^\alpha},
    \end{align}
\end{subequations}
where $G_0=e^2/h$ and $K_0=\pi^2k_{\rm B}^2T/(3h)$ are the quanta of electrical and heat conductance, respectively.
Since both $\mathtt{T}_{\rm AR}^\alpha$ and $\mathtt{T}_{\rm CAR}^\alpha$ are symmetric functions of $\varepsilon$ \cite{draft_Cao2015}, we have $L_{12}^\alpha = L_{21}^\alpha$ and these coefficients only depend on $\mathtt{T}_{\rm EC}^\alpha$. 
Therefore, we also have $\Pi^\alpha=TS^\alpha$, which is why we only have to discuss the Seebeck effect in addition to electrical and heat conductance.

\section{Density of States and Inverted Andreev Bands}
Having established the mathematical framework, in this section we will analyze the ABSs and the transmission functions. 
While the ABSs and $\mathtt{T}_{\rm AR}$ will mainly recover known results, the EC transmission will reveal the main observation of this study, the inversion of Andreev bands.

\subsection{Local density of states and transmission functions}\label{draft sec DOS A}

\begin{figure*}
    \centering
    \includegraphics[width=\linewidth]{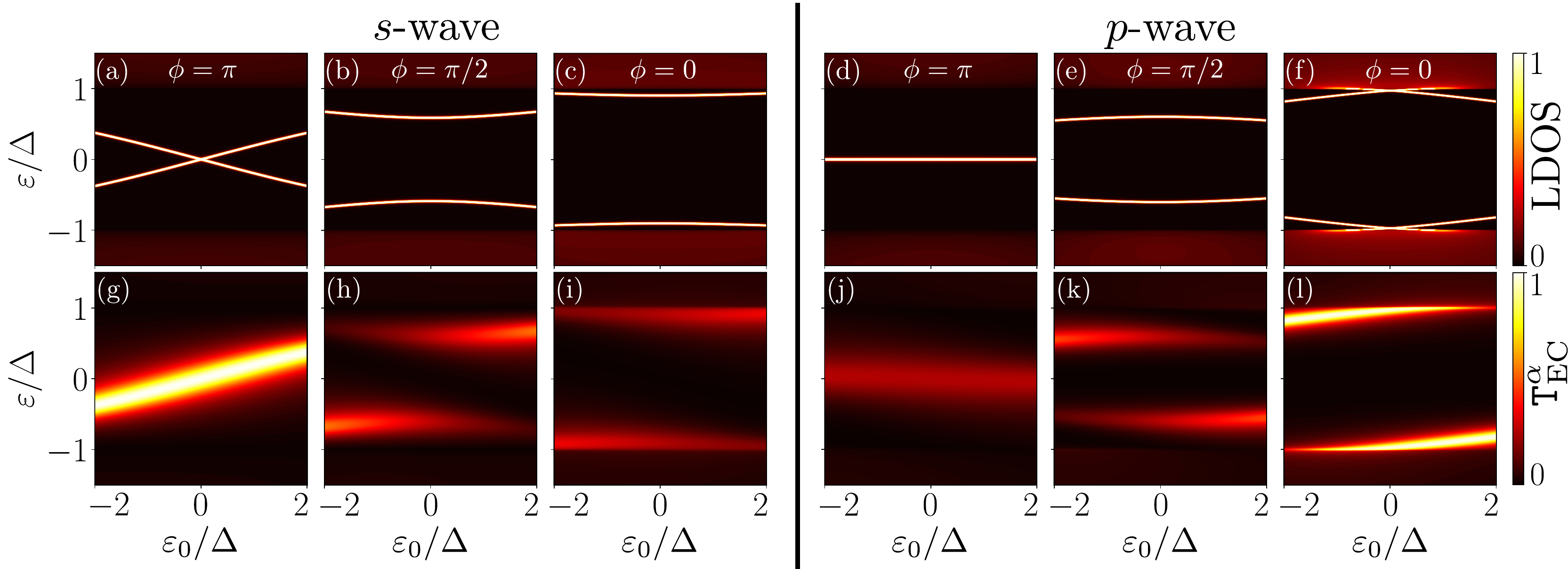}
    \caption{First row: LDOS on the QD, cf. Eq.~\eqref{eq draft LDOS}, as a function of $\varepsilon_0$ and $\varepsilon$.
    We used $\Gamma_1=\Gamma_2=2\Delta$ and $\Gamma_{\rm L}=\Gamma_{\rm R}=0$.
    Second row: $\mathtt{T}_{\rm EC}^\alpha$ as a function of $\varepsilon_0$ and $\varepsilon$.
    Here we used $\Gamma_1=\Gamma_2=2\Delta$ and $\Gamma_{\rm L}=\Gamma_{\rm R}=\Delta/2$.
    In all plots we used the Dynes parameter $\eta=0.001\Delta$.   }
    \label{fig draft 3}
\end{figure*}

We start by analyzing the ABSs forming on the QD. 
For that, we numerically compute the local density of states (LDOS) from the GF as
\begin{equation}\label{eq draft LDOS}
    \mathrm
{LDOS} = -\frac{1}{\pi}\mathrm{Im}\left( G_0^{\alpha, \rm ee} + G_0^{\alpha, \rm hh} \right) \; .
\end{equation}
The results are shown in Figs.~\ref{fig draft 2} (a)-(c) for the case $\alpha=s$ and (d)-(f) for $\alpha=p$ for different values of the QD energy level $\varepsilon_0$.
Notice that the dot level here takes the role of a scattering potential in the continuum approach \cite{draft_BTK1982}:
For $\varepsilon_0=0$, the ABS gap closes at $\phi=\pi$ (a); increasing the dot level lifts the degeneracy and the gap opens (b)-(c). 
In contrast, for $\alpha=p$ the degeneracy at $\phi=\pi$ is conserved for any value of $\varepsilon_0$ (d)-(f) and a gap to the continuum opens.
This protected zero-energy crossing is the most well-known feature of topological JJs and is often referred to as Majorana bound states.

Now we can study the transmission functions $\mathtt{T}_{\rm EC}^\alpha$ and $\mathtt{T}_{\rm AR}^\alpha$, which are defined in Eqs.~\eqref{eq draft transmission} (a)-(b).
We use $\Gamma_{\rm L}=\Gamma_{\rm R}$, implying $\mathtt{T}_{\rm AR}^\alpha=\mathtt{T}_{\rm CAR}^\alpha$, making it sufficient to discuss $\mathtt{T}_{\rm AR}^\alpha$.
Note that small deviations from $\Gamma_{\rm L}=\Gamma_{\rm R}$ and $\Gamma_1=\Gamma_2$ have no significant impact on the observations made below.
We start with the analysis of $\mathtt{T}_{\rm AR}$, which is shown in Figs.~\ref{fig draft 2}~(g)-(i) for $\alpha=s$ and in (j)-(l) for $\alpha=p$.
The transmission is basically given by a broadening of the LDOS with some important details.
Due to particle-hole symmetry, $\mathtt{T}_{\rm AR}$ is a symmetric function of $\varepsilon$ with a constant value of $0.25$ along the LDOS with exception of the gap vanishing points $\phi=\pi$ for $\alpha=s$ and $\phi=0$ for $\alpha=p$, where there is no particle-hole mixing and, hence, no Andreev reflection. 
This can be used for junction characterization:
Since at $\phi=\pi$ the AR transmission vanishes in the $s$-wave case, there will also be zero electrical conductance coming from AR, while for $\alpha=p$ the value will be finite and also quantized as we will show below.

In contrast to $\mathtt{T}_{\rm AR}$, the EC transmission is only finite for states that have a finite electronic component. 
For $\alpha=s$, there is a maximum of transmission around $\phi=\pi$, which is found at positive energies $\varepsilon>0$ if $\varepsilon_0>0$; cf. panels (h) and (i).
The reason for the location of the maximum lies in the induced gap $|\Delta_{\rm ind}^s(\phi)|$, which vanishes at $\phi=\pi$ (if $\Gamma_1=\Gamma_2$) and, hence, there is no particle-hole mixing, leading to a maximal EC transmission. 
A similar feature can be observed for $\alpha=p$: 
The induced gap $|\Delta_{\rm ind}^p(\phi)|$ vanishes at $\phi=0$, which is why the maximal transmission is found at this point.
However, there is a severe difference to the $s$-wave case: 
For finite QD energy $\varepsilon_0>0$, the maximal transmission is found at negative energies. 
This reversed behavior of the EC transmission as a function of energy is a first hint towards the band inversion of the ABSs, which is the main observation of this paper and will be discussed in more detail in the next section.

\subsection{Inverted Andreev bands}

As we have seen above, the transmission $\mathtt{T}_{\rm EC}^p$ shows an anomalous behavior as it has its maximum at negative energies when the dot level is positive. 
To make this more visible and give it a plausible interpretation, we take a different perspective by revisiting the LDOS and EC transmissions as a function of $\varepsilon_0$ for a fixed phase, which yields the plots in Fig.~\ref{fig draft 3}. 

We start again by analyzing the LDOS shown in Figs.\ref{fig draft 3} (a)-(c) for $\alpha=s$ and in (d)-(f) for $\alpha=p$.
In the $s$-wave case, if $\phi=\pi$, the induced gap $|\Delta_{\rm ind}^s(\phi)|$ vanishes and, hence, the dispersion in panel (a) resembles the bare dot dispersion renormalized by the presence of the quasiparticle continuum at $\varepsilon>\Delta$.
As soon as we tune the phase away from $\pi$ in (b) and (c), the gap opens leading to a particle-hole mixing on the dot; cf. the analytical quasiparticle spectrum in Fig.~\ref{fig draft 1}~(c).
All in all, the dispersion on the dot for $\alpha=s$ is close to the usual quasiparticle dispersion in an $s$-wave SC; cf. Fig.~\ref{fig draft 1}~(c).

This is very different in the $p$-wave case.
The zero-energy crossing at $\phi=\pi$ now shows as a constant twofold degenerate zero-energy mode. 
Tuning the phase difference away from $\pi$ lifts the degeneracy (panels (e) and (f)).
Remarkably, the modes have opposite curvature compared to the $s$-wave case.

Let us now analyze again the EC transmission. 
Reading $\mathtt{T}_{\rm EC}^\alpha$ now as a function of $\varepsilon_0$ allows for an easy distinction of electron-like quasiparticles from hole-like ones:
Recall that $\mathtt{T}_{\rm EC}^\alpha$ is only computed from the electronic GF. 
Therefore, the larger the EC transmission the stronger the electronic character of the quasiparticles.
This idea is confirmed in Fig.~\ref{fig draft 3} (g).
Since the induced gap vanishes in this plot, the EC transmission up to the aforementioned renormalization due to the above-gap continuum recovers the usual electron transmission through a resonant level (see, e.g., Ref.~\cite{draft_Cuevas}).
By tuning the phase difference away from $\pi$ in panels (h) and (i), the induced pairing mixes particles and holes on the dot, whereby the electron-like states are found where the signs of $\varepsilon$ and $\varepsilon_0$ match [cf. the red curves in Fig.~\ref{fig draft 1}~(c)].

Comparing these findings to the $p$-wave case shown in panes (j)-(l), one could think at first that here the hole transmission is shown instead of the electrons:
The stripe of finite transmission in panel (j) shows a slight but visible negative slope, which reminds one of the hole transmission in an RTM. 
Moreover, in panels (k) and (l) the states of high transmission are found where the sign of $\varepsilon$ is opposed to the sign of $\varepsilon_0$, which is where one usually would expect the hole states to be.
In summary, we conclude that in the $p$-wave junction electron- and hole-like ABSs have switched places and we call these inverted ABSs.

The band inversion can also be found analytically. 
For this purpose, we consider the self-energies from Eq.~\eqref{eq draft self-energies} in the low-energy limit $\varepsilon/\Delta\ll1$:
\begin{subequations}\label{eq draft self low}
\begin{align}
    \Sigma_{\rm S}^s & \approx 
    \begin{pmatrix}
        0 & \tilde{\Gamma} (\eu^{\imi \phi} + 1) \\
        \tilde{\Gamma} (\eu^{-\imi \phi} + 1) & 0
    \end{pmatrix}  , \\
    \Sigma_{\rm S}^p & \approx -\frac{\Delta}{\varepsilon} 
    \begin{pmatrix}
        -2 \tilde{\Gamma} & \tilde{\Gamma} (\eu^{\imi \phi} - 1) \\
        \tilde{\Gamma} (\eu^{-\imi \phi} - 1) & -2 \tilde{\Gamma}
    \end{pmatrix}  ,
\end{align}
\end{subequations}
where we used gauge freedom to set $\varphi_1=\phi$ and $\varphi_2=0$.
With that we find the low-energy expressions for the EC transmission functions to be (see Appendix~\ref{sec draft app C})
\begin{subequations}
    \begin{align}
        \mathtt{T}_{\rm EC}^s(\varepsilon) & \approx \frac{(\varepsilon+\varepsilon_0)^2 + 4\Gamma^2}{\mathscr{D}^s(\varepsilon^2, \varepsilon_0^2, \phi)}, \\
        \mathtt{T}_{\rm EC}^p(\varepsilon) 
        & \approx \frac{(\varepsilon^2 + \varepsilon\varepsilon_0 - 2\Tilde{\Gamma}\Delta)^2 + 4\varepsilon^2\Gamma^2}{\mathscr{D}^p(\varepsilon^2, \varepsilon_0^2, \phi)},
    \end{align}
\end{subequations}
where we set $\Gamma_{\rm L} = \Gamma_{\rm R} \equiv \Gamma$ and $\Gamma_1 = \Gamma_2 \equiv \Tilde{\Gamma}$.
Since the denominators $\mathscr{D}^\alpha$, which are explicitly given in Appendix~\ref{sec draft app C} in Eq.~\eqref{eq draft denominators}, only depend on $\varepsilon^2$ and $\varepsilon_0^2$, it becomes clear that $\mathtt{T}_{\rm EC}^s$ reaches its maximum when $\varepsilon$ and $\varepsilon_0$ have the same sign.
Contrary to this, $\mathtt{T}_{\rm EC}^p$ takes maximal value when $\varepsilon$ and $\varepsilon_0$ have opposite sign due to the $(\varepsilon^2 + \varepsilon\varepsilon_0 - 2\Tilde{\Gamma}\Delta)^2\approx(\varepsilon\varepsilon_0 - 2\Tilde{\Gamma}\Delta)^2$ term (the $\varepsilon^2$ term is small in the low-energy limit).
As seen from the analytical calculation, this differing behavior in sign originates from the $\Delta/\varepsilon$ term in the induced gap $\Delta_{\rm ind}^p$ given in Eq.~\eqref{eq draft induced p}, which itself comes from the presence of Majorana zero modes at the boundaries of the topological SC wire \cite{draft_zazunov2016}.
The band inversion can be intuitively understood as a consequence of hybridization of Majorana zero modes with the QD energy level $\varepsilon_0$, which we show in Appendix~\ref{sec draft appendix hybridization}.
Therefore, the inversion of Andreev bands is a signature of Majorana zero modes.

\section{Transport Observables}
\begin{figure}
    \centering
    \includegraphics[width=1\linewidth]{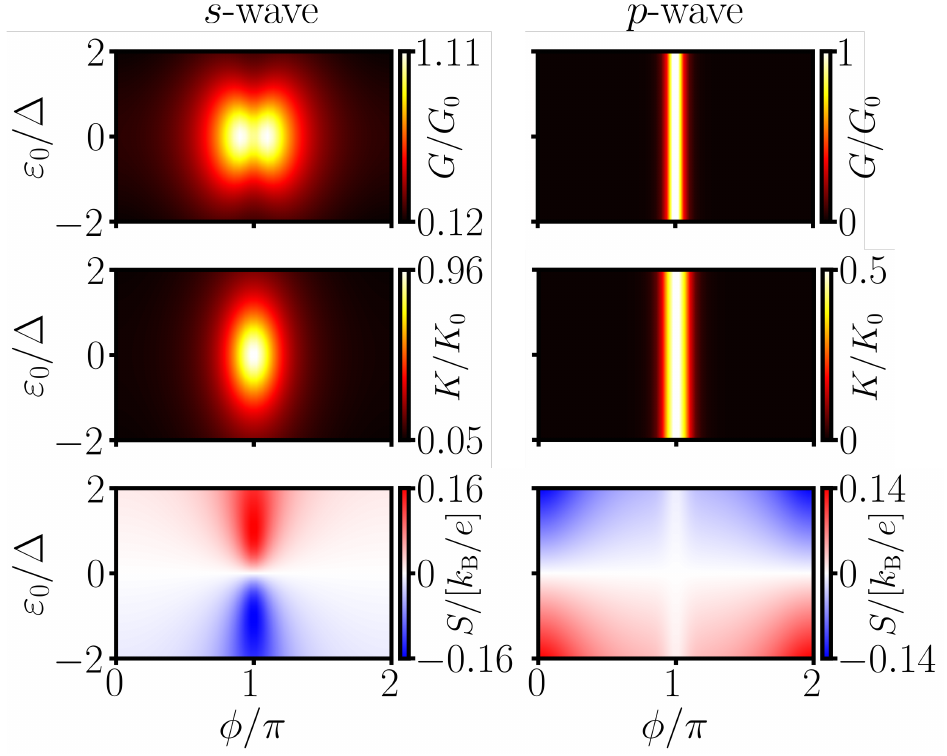}
    \caption{Transport coefficients $G$, $K$, and $S$ as functions of phase difference $\phi$ and dot level $\varepsilon_0$. 
    Parameters: $\Gamma_{1,2}=2\Delta$, $\Gamma_\mathrm{L,R}=0.5\Delta$, $\eta=0.001\Delta$, $k_{\rm B}T=0.01\Delta$, and $\mu=0$.}
    \label{fig draft Figure 4}
\end{figure}
In this section, we turn our attention to the previously introduced transport observables, namely the electrical conductance $G$, the thermal conductance $K$, and the Seebeck coefficient $S$. 
While $G$ and $K$ can be used to identify the pairing symmetry of the superconductors, the anomalous Seebeck effect is a direct consequence of the previously found band inversion. 
Moreover, we will show how the EC conductance $G_{\rm EC}(\mu)$ can be used to measure the band inversion directly.

\subsection{Electrical and heat conductance}
We start by analyzing the electrical conductance $G$, which is shown in the first row in Fig.~\ref{fig draft Figure 4} as a function of phase difference $\phi$ and dot level $\varepsilon_0$.
We use a base temperature of $k_{\rm B}T=0.01\Delta$, justifying the linearized transport coefficients in Eq.~\eqref{eq draft onsager coefficients}.
Higher temperatures generally do not lead to significantly different results as long as the condition $k_{\rm B}T\ll\Delta$ is fulfilled.
There are several remarkable features that allow us to distinguish between the two junctions:
The $s$-wave conductance shows a kidney-like structure with two peaks located at $\varepsilon_0=0$ and slightly away from the symmetry point $\phi=\pi$. 
This is due to the effects of local and crossed Andreev reflection, which both reach their maximal transmission slightly left and right from $\phi=\pi$ while completely vanishing in the center; cf. Fig.~\ref{fig draft 2}~(m).
As discussed above, the behavior of the AR transmission is directly linked to the phase dependence of the induced gap $\Delta_{\rm ind}^s(\phi)$, which also vanishes at $\phi=\pi$. 
Therefore, the kidney structure of the electrical conductance is an indicator of $s$-wave pairing symmetry. 

\begin{figure}
    \centering
    \includegraphics[width=1\linewidth]{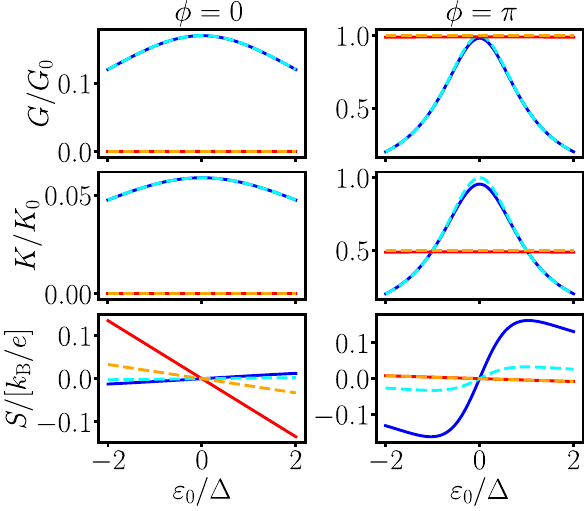}
    \caption{Transport coefficients $G$, $K$, and $S$ for fixed phases $\phi$ as functions of $\varepsilon_0$.
    The blue and red solid curves are the numerical results for $s$- and $p$-wave, respectively, while the cyan and orange dashed curves are computed analytically.
    Parameters: $\Gamma_{1,2}=2\Delta$, $\Gamma_{\rm L,R}=0.5\Delta$, $\eta=0.001\Delta$, $k_{\rm B}T=0.01\Delta$, and $\mu=0$.
    }
    \label{fig draft Figure 5}
\end{figure}

In order to obtain better insight into the functional dependencies of $G^s$, we can analytically compute the conductance in a low-temperature limit by applying a Sommerfeld expansion.
From the result given in Appendix~\ref{sec draft appendix D} in Eq.~\eqref{eq draft approx Gs}, we can read off two special cases with $\phi=0$ and $\phi=\pi$:
\begin{subequations}
    \begin{align}
        G^s(\phi=0)/G_0 & = 4\Gamma^2 \frac{\varepsilon_0^2 + 4\Gamma^2 + 12\Tilde{\Gamma}^2}{ (\varepsilon_0^2 + 4\Gamma^2 + 4\Tilde{\Gamma}^2)^2}, \\
        G^s(\phi=\pi)/G_0 & = \frac{4\Gamma^2}{\varepsilon_0^2 + 4\Gamma^2}.
    \end{align}
\end{subequations}
The first observation from the above equations is that for $\phi=\pi$, the conductance becomes independent of $\Tilde{\Gamma}$, which is the coupling to the superconducting leads.
This is due to the vanishing of the induced gap, making the SCs effectively invisible for the quantum dot. 
What remains is the usual conductance of a resonant tunneling model with normal-lead coupling $\Gamma$, closing the circle to our initial observations made in Fig.~\ref{fig draft 1}~(c) and the discussion thereof. 
As in the resonant tunneling model, the conductance reaches unity at $\varepsilon_0=0$, indicating perfect transmission.
In contrast, for $\phi=0$ the conductance depends on the couplings $\Tilde{\Gamma}$, leading to a significantly suppressed electrical transport. 
The analytical results are shown in the cyan dashed curves in Fig.~\ref{fig draft Figure 5}.
Comparison with the numerical results (blue solid curves) proves that the low-energy limit yields reliable values for the conductance.
The only exception is found around $\phi=\pi$ and $\varepsilon_0=0$, where the numerical values are slightly below unity due to the finite temperature of $k_{\rm B}T=0.01\Delta$.

Also for the $p$-wave case analytical results can be found by using Eq.~\eqref{eq draft approx Gp} in Appendix~\ref{sec draft appendix D}:
\begin{subequations}
    \begin{align}
        G^p(\phi=0)/G_0 & = 0, \\
        G^p(\phi=\pi)/G_0 & = 1.
    \end{align}
\end{subequations}
The quantized value of the conductance for $\phi=\pi$ is a result of the protected zero-energy mode and has been discussed extensively in the literature \cite{draft_Law2009, draft_Flensberg2010, draft_zazunov2016}.
However, also the perfectly vanishing conductance at $\phi=0$ is an indicator of Majorana modes:
First of all, the induced gap $\Delta_{\rm ind}^p$ vanishes at this point, prohibiting AR and CAR processes. 
Moreover, as one can see in the self-energy in Eq.~\eqref{eq draft self-energies b}, the system then behaves like a resonant tunneling model with self-energy 
\begin{equation}
    \Sigma^p(\phi=0) = -2
    \left(
    \Tilde{\Gamma}\sqrt{\Delta^2 - \varepsilon^2} /\varepsilon + \imi\Gamma
    \right)\tau_0.
\end{equation}
The $1/\varepsilon$-term stems from the presence of Majorana modes at the boundaries of the topological superconductors and appears eventually in the denominator $\mathscr{D}^p$ given in Eq.~\eqref{eq draft denominators b}, making the denominator diverge and the conductance vanish.
In summary, while the quantized conductance at $\phi=\pi$ is a footprint of the Majorana bound state in the topological Josephson junction, the vanishing conductance at $\phi=0$ indicates the presence of Majorana modes at the superconductors forming said junction.
Lastly, comparing the orange dashed curve with the red solid curve in Fig.~\ref{fig draft Figure 5} shows a perfect match between analytical and numerical results in the $p$-wave conductance.

Since at low temperatures, both the electrical and heat conductance are proportional to the transmission, the heat transport yields a very similar picture to that of the electrical transport. 
However, since the transmission functions for the two processes differ [there is no contribution from local Andreev reflection to heat transport, see Eq.~\eqref{eq draft transmission onsager}], there are a few differences.
First of all, as we can observe in Fig.~\ref{fig draft Figure 4}, the heat conductance in the $s$-wave case shows no longer the kidney structure that we have seen in the electrical transport, due to the missing Andreev contribution. 
Since the electrical conductance is determined by $G/G_0 \approx \mathtt{T}_{\rm EC} + \mathtt{T}_{\rm CAR} + 2\mathtt{T}_{\rm AR}$ while the heat conductance is given by $K/K_0 \approx \mathtt{T}_{\rm EC} + \mathtt{T}_{\rm CAR}$ [cf. Eq.~\eqref{eq draft transmission onsager}], the comparison between $G$ and $K$ allows us to filter out the Andreev reflection from the transport. 
Since in our case $\mathtt{T}_{\rm CAR} = \mathtt{T}_{\rm AR}$, this also makes it possible to extract the pure EC contribution to the electrical and heat transport. 
This we will use below when we show how the band inversion can be measured using the EC conductance.

The missing AR contribution is also clearly visible in the $p$-wave case:
Although the overall structure of $K^p$ is similar to the electrical conductance, the heat conductance at $\phi=\pi$ is quantized to a value of $K_0/2$ rather than the full heat conductance quantum. 
This is in agreement with the results presented in Ref.~\cite{draft_Bauer2021}, where also a half-quantized heat conductance in a transverse topological Josephson junction has been predicted. 
With our analysis we have shown that the half-quantized value consists of equal parts of EC and CAR contributions.

\subsection{Seebeck coefficient}
While electrical and heat conductance are both determined by the transmission at the chemical potential and, as we have seen, give mostly the same information, the Seebeck coefficient can provide new insights.
Since the Onsager coefficient $L_{12}$ is determined by the derivative of the transmission function (cf. Appendix~\ref{sec draft appendix D}), the thermopower contains information about particle-hole asymmetries in the system.
Therefore, we can expect to see fingerprints of the band inversion in the thermoelectric transport.

The last row of Fig.~\ref{fig draft Figure 4} shows the Seebeck coefficient for $s$- and $p$-wave junctions.
Even though the two pictures look fairly different on first glance, they are actually quite similar except for two major elements:
In comparison to the $s$-wave case, the $p$-wave result appears (a) to be phase shifted by a factor of $\pi$ and (b) to have the opposite sign. 
In short we can observe that $S^p(\phi) \approx -S^s(\phi+\pi)$.

This phase shift we have already observed in the conductances and can again be traced back to the phase shift in the induced gap, which itself is a consequence of the different pairing symmetries in $s$- and $p$-wave superconductors.
The sign change however is a new observation:
Normally, like in the resonant tunneling model, for a positive dot level we would expect the thermoelectric current to be dominated by electrons as they are energetically closer to the energy of the scatterer.
In that case, the Seebeck coefficient would be positive.
Similarly, if $\varepsilon_0<0$, the current would be mainly carried by holes and, therefore, the thermopower would be negative.
This is exactly the picture that we can see in the $s$-wave junction: $\mathrm{sgn}(S^s(\varepsilon_0)) = \mathrm{sgn}(\varepsilon_0)$.

In the $p$-wave junction we observe the opposite behavior, i.e., $\mathrm{sgn}(S^p(\varepsilon_0)) = -\mathrm{sgn}(\varepsilon_0)$.
This indicates that at positive dot levels the states are more hole-like, while the electron-like states are found at negative energies, proving that the Andreev bands are indeed inverted.
In order to get some analytical insight, we can use that at low temperatures we have $S\sim L_{12} \sim \frac{\partial \mathtt{T}_{12}}{\partial\varepsilon}$, giving us (cf. Appendix~\ref{sec draft appendix D})
\begin{subequations}
    \begin{align}
        S^s/[k_{\rm B}/e] & = \frac{\pi^2}{3}k_{\rm B}T \frac{2\varepsilon_0}{\varepsilon_0^2 + 4\Gamma^2 + 12\Tilde{\Gamma}^2\cos^2(\phi/2)}, \\
        S^p/[k_{\rm B}/e] & = \frac{\pi^2}{3} k_{\rm B}T \frac{-\varepsilon_0}{\Tilde{\Gamma}\Delta(1 + 3\sin^2(\phi/2))},
    \end{align}
\end{subequations}
with $\mu=0$ and therefore we have $S^s\propto\varepsilon_0$ and $S^p\propto-\varepsilon_0$.

In the last row of Fig.~\ref{fig draft Figure 5} we compare the analytical and numerical results for $S^\alpha$ at two different phases.
At $\phi=0$, the values for $\alpha=s$ (blue solid and cyan dashed curves) match quite well and show the expected linear behavior $\propto\varepsilon_0$. 
In contrast, the analytical and numerical values for $S^p$ (red solid and orange dashed curves) do not match, although showing the same qualitative behavior.
The opposite pattern can be observed at $\phi=\pi$:
Here, the $p$-wave values match very well, while the $s$-wave curves strongly deviate from each other.

In order to obtain the analytical results for the Seebeck coefficient, we applied two approximations:
the low-energy approximation leading to the simplified self-energies in Eq.~\eqref{eq draft self low} and the Sommerfeld expansion, which can be used at sufficiently low temperatures. 
It is easy to check that lowering the temperature does not lead to a better match between analytical and numerical results, which is why the origin of this mismatch has to lie in the low-energy approximation.
What the low-energy approximation effectively does is to ignore the presence of the quasiparticle continuum above the superconducting gap. 
Since the Seebeck coefficient is mainly determined by the derivative of the transmission function, it also takes energies into account which deviate from 0.
Therefore, unlike the conductances, which only depend on the transmission at zero energy and, therefore, are not influenced by the presence or absence of above-gap quasiparticles, the thermopower can strongly change if the quasiparticles are removed. 
However, as Fig.~\ref{fig draft Figure 5} shows, the analytical results still give the correct qualitative behavior.

\subsection{EC conductance}
\begin{figure}
    \centering
    \includegraphics[width=1\linewidth]{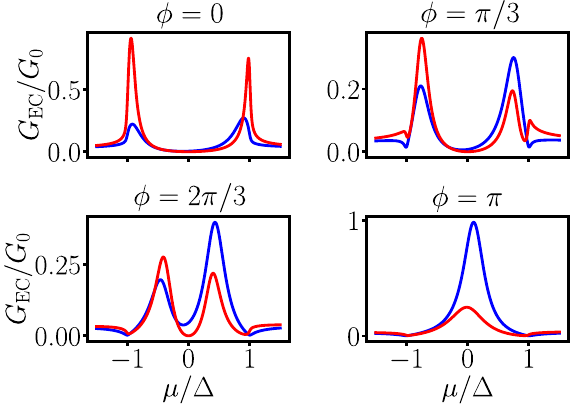}
    \caption{EC conductance $G_{\rm EC}^\alpha$ for different values of $\phi$ and $\varepsilon_0=\Delta/2$. The blue (red) curves correspond to $\alpha=s$ ($\alpha=p$). Other parameters: $\Gamma_{1,2}=2\Delta$, $\Gamma_{\rm L,R}=0.5\Delta$, $\eta=0.001\Delta$, and $k_{\rm B}T=0.01\Delta$.}
    \label{fig draft figure 6}
\end{figure}
A second method to detect the band inversion is to measure the EC conductance as a function of the reference chemical potential $\mu$, whereby EC conductance means the contribution to the conductance coming from EC processes alone.
Since band inversion is clearly visible in the EC transmission in Fig.~\ref{fig draft 3} and the conductance in turn is, at low temperatures, solely determined by the transmission evaluated at the chemical potential, the EC conductance allows us to directly measure the transmission and, therefore, should reveal the band inversion.
As we have argued above, by measuring both the electrical and heat conductance we can approximately determine the contribution of AR processes, which enables the extraction of the more interesting EC conductance $G_{\rm EC}^\alpha$, i.e., the contribution to the conductance originating from EC only.

The numerical results are shown in Fig.~\ref{fig draft figure 6}. 
Note that since the analytical results obtained in Appendix~\ref{sec draft appendix D} used a low-energy approximation, these expressions are only valid for chemical potentials close to $0$. 
Therefore we will only discuss the numerical results here.
As one can see in Fig.~\ref{fig draft figure 6}, for phases $\phi\neq0$ the conductance always has a double-peak structure with one peak higher than the other. 
Since a conductance peak shows us at which energies the electronic states on the quantum dot are located, in a normal resonant tunneling model one would expect a single conductance peak around the dot level $\varepsilon_0$.
However, due to the mixing of particles and holes due to the superconductors, there are also electron-like states at negative energies, although less pronounced. 
Therefore, the higher peak tells us, at what energies we find the electronic states.

In all three plots in Fig.~\ref{fig draft figure 6} which show a double-peak structure, the higher peak for the $s$-wave junction is found at positive chemical potentials, while in the $p$-wave case we find them at negative energies. 
This proves again that in the topological junction the electronic states and the holes have switched their energetic positions and, therefore, the Andreev bands are inverted. 
Another observation that one can make is that the peaks move towards each other when tuning the phase away from $0$ until eventually merging at $\phi=\pi$. 
This is due to the fact that the Andreev bound states also approach (cross) each other in the $s$-wave ($p$-wave) junction, cf. Fig.~\ref{fig draft 2}. 
At $\phi=\pi$ the conductance peak for the $p$-wave case is located at zero energy (where we expect the Majorana bound state to be), while the $s$-wave peak is moved slightly away from $\mu=0$, 
showing that the zero-energy state in the topological junction is protected while the gap is opened in the $s$-wave junction.

\section{Conclusions}
We have studied the transport signatures in a four-terminal junction consisting of two normal metal leads and two superconductors coupled to a central quantum dot.
We were interested in the transverse thermoelectric transport between the normal leads, i.e., transverse to the formed Josephson junction, enabling us to look into the Andreev bound states forming on the dot. 
We have chosen the SC leads to have either $s$-wave or $p$-wave pairing in order to compare topological JJs with conventional ones.
We have shown how the physics of the ABSs can be fully understood from the induced pairing $\Delta_{\rm ind}^\alpha$, which, depending on the pairing symmetry of the superconductors, depends differently on the phase bias.
Evidence of different phase dependencies of the induced gap can be found in the electrical conductance, cf. Fig.~\ref{fig draft Figure 4}.
This yields the possibility of classifying the junction.

We studied the local density of states and observed the well-known protected crossing of the ABSs in the $p$-wave junction.
By looking at the the transmission functions corresponding to transverse cotunneling and Andreev reflection processes, we made two things visible:
The Andreev transmission mirrors well the dependencies of the induced gap function, especially the points where the induced gap vanishes, leading to a vanishing transmission. 
On the other hand, the EC transmission showed that the electron and hole states in the topological junction are interchanged, while the Andreev bands in the $s$-wave case behave quite naturally, cf. Fig.~\ref{fig draft 3}.
In analogy to the band inversion in topological materials (see, e.g., Refs.~\cite{draft_Fu2007, draft_Zhu2012}) we called this phenomenon inverted Andreev bands.

We have found evidence of the band inversion in two observables: the Seebeck coefficient and the EC conductance. 
For the Seebeck coefficient we have found that, in the $p$-wave case, an anomalous sign change occurs as opposed to the $s$-wave case. 
This anomalous Seebeck effect has been found before in QD structures containing Majorana zero modes (see, e.g., Refs.~\cite{draft_Lopez2014, draft_Valentini2015, draft_Grosu2023}), but no explanation has been given so far. 
With the inversion of Andreev bands in topological junctions we have found an intuitive explanation for this effect.
We note that in contrast to Refs.~\cite{draft_Lopez2014, draft_Grosu2023}, our theory does not make use of an effective Majorana Hamiltonian, but works with an actual superconducting junction instead, bringing our study closer to experiments. 
Moreover, the four-terminal geometry allows for phase-dependent measurements, extending the three-terminal approach of Refs.~\cite{draft_Lopez2014, draft_Valentini2015, draft_Grosu2023}.

Thermoelectric measurements were established many years ago \cite{draft_Molenkamp1992} and have been further developed since. 
While for the $s$-wave leads any conventional superconductor can be used, for the $p$-wave case we suggest using nanowires with strong spin-orbit coupling \cite{draft_Oreg2010, draft_Deng2016}.
Regarding recent progress in the fabrication of multi-terminal phase-controlled junctions \cite{draft_Antonelli2025}, we are confident that the setup studied in this work can be realized with state-of-the-art experimental techniques.

The occurrence of the anomalous Seebeck effect in Refs.~\cite{draft_Lopez2014, draft_Grosu2023} involving Majorana zero modes leads to the conclusion that band inversion on the quantum dot is a necessary condition for the presence of Majorana bound states. 
However, other effects like Cooper-pair splitting \cite{draft_Klees2023} or correlations \cite{draft_Chi2020} can undo or mimic the anomalous Seebeck effect as they can alter the hybridization on the QD. 
Therefore, future works might concentrate on the precise conditions for the occurrence of the anomalous Seebeck effect and how it can be utilized (perhaps in combination with other effects) for the unambiguous detection of Majorana zero modes.

\begin{acknowledgments}
We would like to thank Florian Goth for helpful discussions.
We acknowledge funding by the Deutsche
Forschungsgemeinschaft (DFG, German Research Foundation) through SFB1170 ToCoTronics, Project-ID
258499086 and through the Würzburg-Dresden Cluster of
Excellence on Complexity and Topology in Quantum
Matter–\textit{ct.qmat} (EXC2147, Project-ID 390858490).
\end{acknowledgments}

\appendix

\section{Hybridization with Majorana zero mode}
\label{sec draft appendix hybridization}
\begin{figure*}
    \centering
    \includegraphics[width=1\linewidth]{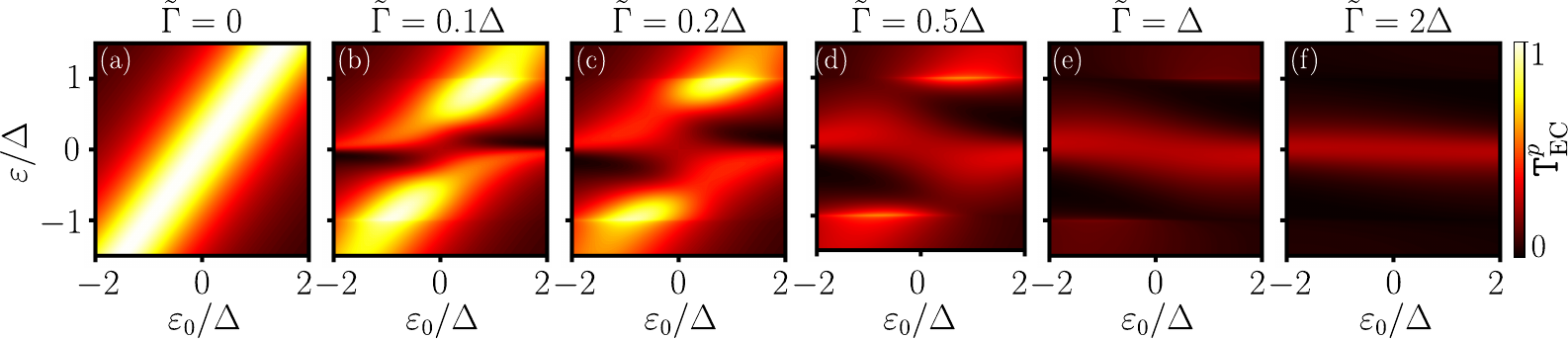}
    \caption{Transmission function $\mathtt{T}_{\rm EC}^p$ as function of dot energy $\varepsilon_0$ and energy $\varepsilon$ at $\phi=\pi$ for various couplings $\Gamma_1=\Gamma_2\equiv\Tilde{\Gamma}$ to the superconductors.
    Other parameters are $\Gamma_{\rm L} = \Gamma_{\rm R}=0.5\Delta$ and $\eta=0.001\Delta$. 
    }
    \label{fig draft figure Appendix}
\end{figure*}
The central observation of this study is that electron and hole bands on the QD are inverted when coupled to $p$-wave superconductors. 
As we will illustrate here, this band inversion originates from a hybridization with the Majorana zero modes localized on the boundaries of the superconductors. 
Similar to the band inversion itself, which is shown in Fig.~\ref{fig draft 3} in the main text, the hybridization is best visible in the EC transmission, $\mathtt{T}_{\rm EC}^p$.

Figure~\ref{fig draft figure Appendix} shows $\mathtt{T}_{\rm EC}^p$ for fixed phase $\phi=\pi$ when increasing the coupling to the superconductors, $\Gamma_1=\Gamma_2\equiv\Tilde{\Gamma}$.
If the superconductors are completely detached ($\Tilde{\Gamma}=0$), the system reduces to a standard resonant tunneling model, for which the transmission is given by \cite{draft_Cuevas}
\begin{equation}\label{eq draft RTM transmission}
    \mathtt{T}_{\rm RTM}(\varepsilon) = \frac{4\Gamma_{\rm L}\Gamma_{\rm R}}{(\varepsilon-\varepsilon_0)^2 + (\Gamma_{\rm L}+\Gamma_{\rm R})^2},
\end{equation}
which is shown in Fig.~\ref{fig draft figure Appendix}~(a).
When starting to slightly couple the superconductors to the QD, as we do in Fig.~\ref{fig draft figure Appendix}~(b), we can see that the single electron band from the first panel hybridizes with the zero-energy mode. 
Due to the hybridization with the Majorana zero mode, the electronic states are pushed toward positive (negative) QD energies $\varepsilon_0$ for negative (positive) $\varepsilon$. 
Increasing the couplings in Figs.~\ref{fig draft figure Appendix}~(c)-(e) further pronounces the band inversion until we reobtain the picture in Fig.~\ref{fig draft 3}~(j) in the last panel in Fig.~\ref{fig draft figure Appendix}.

\section{Derivation of the transmission functions}\label{sec draft app A}
In this appendix, we will show how to derive the exact expressions of the transmission functions given in Eq.~\eqref{eq draft transmission}.
As stated in the main text, we start from the Heisenberg equation of motion to derive an expression for the electrical current:
\begin{subequations}
    \begin{align}
        I_{\rm L}(t) 
        & = \frac{e}{\imi\hbar} \braket{[H, N_{\rm L}]}(t) \\
        & = \frac{e}{\hbar} t_{\rm L} \sum_\sigma
        \left(
        \imi\braket{c_{\rm L\sigma}^\dagger c_{0\sigma}} - \imi\braket{c_{0\sigma}^\dagger c_{\rm L\sigma}}
        \right),
    \end{align}
\end{subequations}
where $N_{\rm L}=\sum_\sigma c_{\rm L\sigma}^\dagger c_{\rm L\sigma}$ is the particle number operator of the left lead and $H$ is the Hamiltonian defined in Eq.~\eqref{eq draft Hamiltonian}.
All operators in the equation above are assumed to be time-dependent in the Heisenberg picture.
By invoking the definition of the lesser Green's function,
\begin{equation}
    G_{ij}^<(t, t') = \imi \braket{\psi_j^\dagger(t') \otimes \psi_i(t)} \; ,
\end{equation}
where $\psi_i^\dagger= (c_{i\uparrow}^\dagger , \ c_{i\downarrow})$ ($i\in \{\mathrm{L,R,1,2,0} \}$) are the Nambu spinors used in the main text, 
we can write the current in Fourier space as 
\begin{equation}\label{eq draft appendix A current}
    I_{\rm L} = \frac{e}{h} t_{\rm L} \int_{-\infty}^\infty \der\varepsilon \ \mathrm{tr} \Big( G_{0\rm L}^<(\varepsilon) - G_{\rm L 0}^<(\varepsilon) \Big) \; .
\end{equation}
Note that we can safely Fourier transform the Green's function since we do not apply any voltage bias to the superconductors, which is why we achieve a steady current without driving the system into a fully non-equilibrium state. 

We can further rewrite the lesser Green's functions by using the Dyson equations \cite{draft_Cuevas}
\begin{subequations}
    \begin{align}
        G_{0\rm L}^< & = G_{00}^{\rm r} V_{0\rm L}g_{\rm LL}^< + G_{00}^< V_{\rm 0L} g_{\rm LL}^{\rm a} \; , \\
        G_{\rm L0}^< & = g_{\rm LL}^< V_{\rm L0} G_{00}^{\rm a} + g_{\rm LL}^{\rm r} V_{\rm L0} G_{00}^< \; .
    \end{align}
\end{subequations}
This now allows us to apply the relations $G_{ii}^{\rm a} - G_{ii}^{\rm r} = G_{ii}^< - G_{ii}^>$ \cite{draft_Cuevas}, which yields
\begin{equation}
    I_{\rm L} = \frac{e}{h} t_{\rm L}^2 \int_{-\infty}^\infty \der\varepsilon \ \mathrm{tr} \Big[ \tau_3 \big( G_{00}^> g_{\rm LL}^< - G_{00}^< g_{\rm LL}^> \big) \Big]  ,
\end{equation}
where $\tau_3$ is the third Pauli matrix and the trace $\mathrm{tr}$ runs over the particle-hole degrees of freedom.
In order to replace the dressed greater and lesser Green's functions with the easier to handle retarded and advanced Green's functions, we use the Dyson equation \cite{draft_Cuevas}
\begin{equation}
    G_{00}^{<,>} = \sum_{i=\rm L,R,1,2,0} \left( 1 + G^{\rm r} V \right)_{0i} g_{ii}^{<,>} \left( 1 + V G^{\rm a} \right)_{i0} \; .
\end{equation}
Recalling that the Green's functions $g_{ii}^{<,>}$ are proportional to the equilibrium local density of states in site $i$ and since we are only interested in the subgap transport, we can neglect the terms with $i=1,2$. 
Furthermore, one can show that also the term proportional to $g_{00}^{<,>}$ vanishes \cite{draft_Cuevas}, which leaves us only two terms to evaluate.
Using that $g_{ii}^< = 2\pi\imi \nu_0 \mathrm{diag}(f_i^{\rm e}, \ f_i^{\rm h})$ and $g_{ii}^> = -2\pi\imi\nu_0 (1- \mathrm{diag}(f_i^{\rm e}, \ f_i^{\rm h}))$ for $i={\rm L,R}$, one finds after some algebra
\begin{subequations}\label{eq draft Landauer}
\begin{align}
    I_{\rm L} = \frac{2e}{h} 4\Gamma_{\rm L} \int_{-\infty}^\infty \der\varepsilon\Big[ & \hspace{0.4cm} \Gamma_{\rm L} \Big| G_{00}^{\rm eh} \Big|^2 \hspace{0.1cm}
    \Big( f_{\rm L}^{\rm e} - f_{\rm L}^{\rm h}\Big)      \\
    & + \Gamma_{\rm R} \Big| G_{00}^{\rm ee} \Big|^2 \hspace{0.1cm} 
    \Big( f_{\rm L}^{\rm e} - f_{\rm R}^{\rm e}  \Big)  \\
    & + \Gamma_{\rm R}\Big| G_{00}^{\rm eh}\Big|^2  \hspace{0.1cm}
    \Big( f_{\rm L}^{\rm e} - f_{\rm R}^{\rm h}  \Big)\Big]  .
\end{align}
\end{subequations}
From this Landauer-type formula we can now read out the expressions for the transmission functions given in Eq.~\eqref{eq draft transmission}.
Note that in this derivation we explicitly took into account the spin, leading to the prefactor of $2$ in the Landauer formula, which is left out in the spinless case.

\section{Derivation of the Onsager coefficients}\label{sec draft app B}
With the Landauer formula found in Eq.~\eqref{eq draft Landauer}, we can now derive the exact expression of the Onsager coefficients given in Eq.~\eqref{eq draft onsager coefficients}.
We start by noting that one can derive a similar Landauer formula to Eq.~\eqref{eq draft Landauer} for the heat current by starting from the equation of motion
\begin{equation}
    J_{\rm L}(t) = -\frac{1}{\imi\hbar} \braket{[H, H_{\rm L} - \mu_{\rm L}N_{\rm L}]}(t)
\end{equation}
and following the same steps as above for the electrical current, leading to 
\begin{subequations}\label{eq draft Landauer heat}
\begin{align}
    J_{\rm L} = \frac{1}{h} 
    \int_{-\infty}^\infty \der\varepsilon (\varepsilon-\mu_{\rm L}) 
    \Big[ & \hspace{0.4cm}\mathtt{T}_{\rm AR}^\alpha \hspace{0.2cm}
    \Big( f_{\rm L}^{\rm e} - f_{\rm L}^{\rm h}\Big)      \\
    & + \mathtt{T}_{\rm EC}^\alpha \hspace{0.2cm} 
    \Big( f_{\rm L}^{\rm e} - f_{\rm R}^{\rm e}  \Big)  \\
    & + \mathtt{T}_{\rm CAR}^\alpha  \hspace{0.0cm}
    \Big( f_{\rm L}^{\rm e} - f_{\rm R}^{\rm h}  \Big)\Big]  .
\end{align}
\end{subequations}
The Fermi functions are given by 
\begin{equation}
    f_i^{\rm e/h}(\varepsilon) = 
    \left(
    1 + \exp
    \left(
    \frac{\varepsilon \mp \mu_i}{k_{\rm B}T_i}
    \right)
    \right)^{-1},
\end{equation}
with $i\in\{\mathrm{L,R}\}$.
By choosing $\mu_{\rm L} = \mu + \Delta\mu$ and $\mu_{\rm R} = \mu$ as well as $T_{\rm L} = T + \Delta T$ and $T_{\rm R} = T$ in accordance with the main text, we can linearize the Landauer formulas assuming $\Delta\mu$ and $\Delta T$ to be small.
This eventually leads to the linear Onsager relation in Eq.~\eqref{eq draft onsager} with the coefficients given in Eq.~\eqref{eq draft onsager coefficients}.

\section{Low-energy expressions of the transmissions}\label{sec draft app C}
To find the low-energy limits of the transmission functions we start by noting that if $\varepsilon/\Delta\ll1$, we have 
\begin{equation}
    \frac{\Delta}{\sqrt{\Delta^2-\varepsilon^2}}\approx 1, \ \ \ \Tilde{\varepsilon}=\frac{\varepsilon}{\sqrt{\Delta^2-\varepsilon^2}}\approx0,
\end{equation}
leading to the self-energies in Eq.~\eqref{eq draft self low}.
In the $s$-wave case, the Green's function is then given by (cf. Eq.~\eqref{eq draft dressed GF})
\begin{equation}
    G_0^s \approx 
    \begin{pmatrix}
        \varepsilon - \varepsilon_0 + 2\imi\Gamma 
        & -\Tilde{\Gamma}(\eu^{\imi\phi} + 1) \\
        -\Tilde{\Gamma}(\eu^{-\imi\phi} + 1) 
        & \varepsilon + \varepsilon_0 + 2\imi\Gamma
    \end{pmatrix}^{-1}
\end{equation}
By explicitly calculating the inverse and using definition in Eq.~\eqref{eq draft transmission}, we find
\begin{subequations}\label{eq draft approx T s}
    \begin{align}
        \mathtt{T}_{\rm EC}^s = 4\Gamma^2 |G_0^{s,\rm ee}|^2 
        & = \frac{(\varepsilon+\varepsilon_0)^2 + 4\Gamma^2}{\mathscr{D}^s(\varepsilon^2, \varepsilon_0^2, \phi)}, \label{eq draft approx T s EC} \\
        \mathtt{T}_{\rm AR}^s = 4\Gamma^2 |G_0^{s, \rm eh}|^2
        & = \frac{4\Tilde{\Gamma}^2\cos^2(\phi/2)}{\mathscr{D}^s(\varepsilon^2, \varepsilon_0^2, \phi)}.
    \end{align}
\end{subequations}
In the same way we can calculate the transmissions in the $p$-wave case, which are given by
\begin{subequations}\label{eq draft approx T p}
    \begin{align}
        \mathtt{T}_{\rm EC}^p 
        & = \frac{(\varepsilon^2 + \varepsilon\varepsilon_0 - 2\Tilde{\Gamma}\Delta)^2 + 4\varepsilon^2\Gamma^2}{\mathscr{D}^p(\varepsilon^2, \varepsilon_0^2, \phi)} , \label{eq draft approx T p EC} \\
        \mathtt{T}_{\rm AR}^p 
        & = \frac{4\Delta^2\Tilde{\Gamma}^2\sin^2(\phi/2)}{\mathscr{D}^p(\varepsilon^2, \varepsilon_0^2, \phi)},
    \end{align}
\end{subequations}
where the denominators are given by
\begin{widetext}
\begin{subequations}\label{eq draft denominators}
    \begin{align}
        \mathscr{D}^s(\varepsilon^2, \varepsilon_0^2, \phi) & = \frac{1}{4\Gamma^2}
        \left(
        \left(\varepsilon^2 - \varepsilon_0^2 - 4\Gamma^2 - 4\Tilde{\Gamma}^2\cos^2(\phi/2)
        \right)^2 + 16\varepsilon^2\Gamma^2
        \right), \\
        \label{eq draft denominators b}
        \mathscr{D}^p(\varepsilon^2, \varepsilon_0^2, \phi)
        & = \frac{\varepsilon^2}{4\Gamma^2}
        \left(
        \left(m(\varepsilon)^2 - 4\Gamma^2 - \varepsilon_0^2 - 4\Delta^2\Tilde{\Gamma}^2\sin^2(\phi/2)/\varepsilon^2
        \right)^2 
        + 16\Gamma^2m(\varepsilon)^2
        \right),
    \end{align}
\end{subequations}
\end{widetext}
with $m(\varepsilon) = \varepsilon - 2\Delta\Tilde{\Gamma}/\varepsilon$.
Note that in the limit $\Tilde{\Gamma}\to0$ (detaching the superconductors), the transmissions in Eqs.~\eqref{eq draft approx T s EC} and \eqref{eq draft approx T p EC} properly reduce to the transmission of the resonant tunneling model in Eq.~\eqref{eq draft RTM transmission}.

\section{Low-temperature expansion of the transport coefficients}\label{sec draft appendix D}
At sufficiently low temperatures, we can approximate the Onsager coefficients given in Eq.~\eqref{eq draft onsager coefficients} by means of a Sommerfeld expansion, resulting in
\begin{subequations}
    \begin{align}
        L_{11} & \approx \mathtt{T}_{11}(\mu), \\
        L_{12} & \approx \frac{\pi^2}{3}k_{\rm B}T \frac{\partial \mathtt{T_{12}}}{\partial\varepsilon}(\mu) = L_{21}, \\
        L_{22} & \approx \frac{\pi^2}{3}\mathtt{T}_{22}(\mu).
    \end{align}
\end{subequations}
The derivative of the transmission function needed for the $L_{12}$ coefficient is given by
\begin{subequations}
    \begin{align}
        \frac{\partial\mathtt{T}_{12}^s}{\partial\varepsilon}\Big|_{\varepsilon=0} 
        & = \frac{2\varepsilon_0}{\mathscr{D}^s(0)}, \\
        \frac{\partial\mathtt{T}_{12}^p}{\partial\varepsilon}\Big|_{\varepsilon=0} 
        & = - \frac{4\Tilde{\Gamma}\Delta\varepsilon_0}{\mathscr{D}^p(0)},
    \end{align}
\end{subequations}
where we used that $\mathtt{T}_{12}^\alpha = \mathtt{T}_{\rm EC}^\alpha + \mathtt{T}_{\rm CAR}^\alpha$.
\begin{widetext}
With that, we find the analytical expressions for the transport coefficients in the $s$-wave case to be
\begin{subequations}
    \begin{align}
        G^s/G_0 & = \frac{(\mu+\varepsilon_0)^2 + 4\Gamma^2 + 12\Tilde{\Gamma}^2\cos^2(\phi/2)}{\mathscr{D}^s(\mu^2, \varepsilon_0^2, \phi)},  \label{eq draft approx Gs} \\
        S^s/[k_{\rm B}/e] & = \frac{\pi^2}{3}k_{\rm B}T \frac{2\varepsilon_0}{\varepsilon_0^2 + 4\Gamma^2 + 3\cdot 4\Tilde{\Gamma}^2\cos^2(\phi/2)} \ \ \ (\mu=0), \label{eq draft approx Ss} \\
        K^s/K_0 & \approx \frac{(\mu+\varepsilon_0)^2 + 4\Gamma^2 + 4\Tilde{\Gamma}^2\cos^2(\phi/2)}{\mathscr{D}^s(\mu^2, \varepsilon_0^2, \phi)}, \label{eq draft approx Ks}
    \end{align}
\end{subequations}
where for the heat conductance we used that the term $L_{12}^2/L_{11}$ is small compared to $L_{22}$ for sufficiently low temperatures. 
Note that for the Seebeck coefficient we put $\mu=0$ for simplicity.
For the $p$-wave case we find
\begin{subequations}
    \begin{align}
        G^p/G_0 & = \frac{(\mu^2+\mu\varepsilon_0-2\Tilde{\Gamma}\Delta)^2 + 4\mu^2\Gamma^2 + 12\Delta^2\Tilde{\Gamma}^2\sin^2(\phi/2)}{\mathscr{D}^p(\mu^2,\varepsilon_0^2,\phi)}, \label{eq draft approx Gp} \\
        S^p/[k_{\rm B}/e] 
        & = - \frac{\pi^2}{3}k_{\rm B}T \frac{\varepsilon_0}{\Tilde{\Gamma}\Delta( 1 + 3\sin^2(\phi/2))} \ \ \ (\mu=0), \label{eq draft approx Sp} \\
        K^p/K_0 & \approx \frac{(\mu^2+\mu\varepsilon_0-2\Tilde{\Gamma}\Delta)^2 + 4\mu^2\Gamma^2 + 4\Delta^2\Tilde{\Gamma}^2\sin^2(\phi/2)}{\mathscr{D}^p(\mu^2,\varepsilon_0^2,\phi)}. \label{eq draft approx Kp}
    \end{align}
\end{subequations}
\end{widetext}

\bibliography{bib.bib}

\end{document}